\documentclass[aps,twocolumn,showpacs,A4,floatfix]{revtex4}
\usepackage{amsmath}
\usepackage{amssymb}
\usepackage{graphicx}
\usepackage{latexsym}

\newcommand{\half} {\frac{1}{2}}
\newcommand{\e} { {\rm e} }
\newcommand{\dv} { {\rm d} }
\newcommand{\iv} { {\it v} \phi_b^2 }

\newcommand{\yy} {\zeta}
\newcommand {\ys} {\left|\yy_s\right|}

\newcommand {\scharge} {\left|\sigma\right|}
\newcommand {\cs} {c_{\rm salt} }
\newcommand {\eps} {\varepsilon_{\rm ps}}
\begin{document}

\title{ Phase Behavior of Polyelectrolyte-Surfactant Complexes
    at Planar Surfaces\\}

\author {Adi Shafir}
\email{shafira@post.tau.ac.il}
\author {David Andelman}
\email{andelman@post.tau.ac.il}
\affiliation{School of Physics and Astronomy   \\
    Raymond and Beverly Sackler Faculty of Exact Sciences \\
    Tel Aviv University, Ramat Aviv, Tel Aviv 69978, Israel\\}

\bigskip
\date{07 August 2006}
\bigskip
\vskip 1truecm

%------------------------------------------------------------
% Abstract
%------------------------------------------------------------
\begin{abstract}

We investigate theoretically the phase diagram of an insoluble
charged surfactant monolayer in contact with a semi-dilute
polyelectrolyte solution (of opposite charge). The polyelectrolytes
are assumed to have long-range and attractive (electrostatic)
interaction with the surfactant molecules. In addition, we introduce
a short-range (chemical) interaction which is either attractive or
repulsive. The surfactant monolayer can have a lateral phase
separation between dilute and condensed phases. Three different
regimes of the coupled system are investigated depending on system
parameters. A regime where the polyelectrolyte is depleted due to
short range repulsion from the surface, and two adsorption regimes,
one being dominated by electrostatics, whereas the other by short
range chemical attraction (similar to neutral polymers). When the
polyelectrolyte is more attracted (or at least less repelled) by the
surfactant molecules as compared with the bare water/air interface,
it will shift upwards the surfactant critical temperature. For
repulsive short-range interactions the effect is opposite. Finally,
the addition of salt to the solution is found to increase the
critical temperature for attractive surfaces, but does not show any
significant effect for repulsive surfaces.

\end{abstract}

\maketitle

%%%%%%%%%%%%%%%%%%%%%%%%%%%%%%%%%%%%%%%%%%%%%%%%%%%%%%%%%%%%%%%%%%%%%%%%%%%%%%
%%%%%%%%%%%%%%%%%%%%%%%%%%%%%%%%%%%%%%%%%%%%%%%%%%%%%%%%%%%%%%%%%%%%%%%%%%%%%%
\section{Introduction}
\label{Intro} 
%%%%%%%%%%%%%%%%%%%%%%%%%%%%%%%%%%%%%%%%%%%%%%%%%%%%%%%%%%%%%%%%%%%%%%%%%%%%%%
%%%%%%%%%%%%%%%%%%%%%%%%%%%%%%%%%%%%%%%%%%%%%%%%%%%%%%%%%%%%%%%%%%%%%%%%%%%%%%

Complexation between surfactants and macromolecules has generated
a great deal of research, both theoretically
\cite{deGennes,andelmanjoanny,chatellier,harries1,mbamala,harries2,may,tzlil,tzlil2,roux,gonzalez,may1,bieshuvel,netz}
and experimentally
\cite{leonenko1,leonenko2,leonenko3,maier1,wu,Ianoul,merkle,monteux1,monteux2}.
This research is motivated in part by the role
surfactant-macromolecule complexes plays in biological
systems~\cite{tzlil,roux}, where the adsorption of a variety of
macromolecules onto charged lipid cell membranes is a precursor
to many important biological processes.

The complexation of charged surfactants and macromolecules, besides
its relevance to biological physics, carries interesting questions
about the interplay of short-range interactions between surfactants
and macromolecules, as compared with long-range electrostatic
interactions. We consider a multicomponent monolayer where the
possibility of  a lateral phase demixing between the various
constitutes exists and is characterized by a demixing (miscibility)
curve in the temperature-concentration plane. For very long
polyelectrolyte (PE) chains, the electrostatic interactions have two
opposite effects on the adsorption. Electrostatic attraction between
surfactants and PE chains increases the complexation, while the
electrostatic repulsion between the PE monomers tends to decrease
the complexation. This interplay also stimulated the interest in the
adsorption of polyelectrolytes on oppositely charged surfaces
~\cite{wiegel,muthu,chatj,varoqui2,varoqui1,joanny,itamar1,itamar2,itamar3,us1,us2,wang1,borisov,dobrynin}.

Several authors have addressed a closely related problem of neutral
polymers adsorbing on a surfactant
monolayer~\cite{chatellier,tzlil,roux}. It was shown that for
polymer-surfactant systems the surfactant demixing occurs for higher
temperatures as compared to a pure surfactant system. Other works
dealt with another similar problem of globular proteins (rigid
macromolecules with no flexibility) adsorbed on charged
membranes~\cite{harries1,harries2}.  A recent paper~\cite{tzlil2}
also deals with the adsorption of one flexible polyelectrolyte chain
on a surfactant monolayer using Monte Carlo simulations. The
possibility that the polyelectrolytes can desorb entirely from the
charged surface when the solution has sufficiently strong ionic
strength~\cite{us1} was not previously discussed.

In this paper, we analyze the demixing of surfactants on the
monolayer plane using numerical solutions of the mean-field
equations. We find that the short-range interactions between the PE
and the surfactants  have a greater effect on the shape of the phase
separation curve than the long-range electrostatics. Three regimes
for the surfactant-polyelectrolyte complexes are found: (i) a
depletion regime, where the polyelectrolytes do not adsorb at all
onto the surfactant layer; (ii) an electrostatic adsorption regime,
where the main attractive interaction between surfactants and
polyelectrolytes is the electrostatic attraction; and (iii) a
chemical adsorption regime, where the main interaction between the
polyelectrolytes and the surface is short-ranged (chemical) and
attractive. In the first two regimes, the surfactant demixing curve
(binodal line) decreases in comparison to that of the pure
surfactant monolayer, while in the third regime  the demixing curve
increases. We also find that for the electrostatic adsorption
regime, the adsorbed amount for a system in the coexistence region
is almost the same as a monophasic system with the same average
concentration. In the chemical adsorption case the adsorbed amount
increases more substantially due to the phase demixing.

The outline of this paper is as follows: we start by introducing the
free-energy functional for a PE solution with salt in contact with a
charged surfactant monolayer. In Sec.~\ref{phasedi} we present our
numerical results and discussion. In Sec.~\ref{analitical} we
present a simple scaling analysis for three limits: (i) PE depletion
from a short-range repulsive surface; (ii) PE electrostatic
adsorption to a short-range repulsive surface; and (iii) the
short-range attractive surface case. Finally, we present some
concluding remarks in Sec.~\ref{conclusion}.

%%%%%%%%%%%%%%%%%%%%%%%%%%%%%%%%%%%%%%%%%%%%%%%%%%%%%%%%%%%%%%%%%%%%%%%%%%%%%%
%%%%%%%%%%%%%%%%%%%%%%%%%%%%%%%%%%%%%%%%%%%%%%%%%%%%%%%%%%%%%%%%%%%%%%%%%%%%%%
\section{Free Energy Formulation}
\label{freeenergy}
%%%%%%%%%%%%%%%%%%%%%%%%%%%%%%%%%%%%%%%%%%%%%%%%%%%%%%%%%%%%%%%%%%%%%%%%%%%%%%
%%%%%%%%%%%%%%%%%%%%%%%%%%%%%%%%%%%%%%%%%%%%%%%%%%%%%%%%%%%%%%%%%%%%%%%%%%%%%%

Consider an aqueous solution of polyelectrolyte chains and salt
ions. The bulk PE solution is in contact with a planar and rigid
interface, on which resides an insoluble monolayer of charged
surfactants. This monolayer can be thought of as a Langmuir
monolayer at the air-water interface, or as the outer leaflet of a
lipid bilayer membrane. The charged surfactants are assumed to be
able to move freely on the interface, but can not dissolve into the
solution (see Figure 1).

The free energy of such a system is composed of three contributions.
The monolayer free energy $F_{\rm surf}$, the solution free energy
$F_{\rm sol}$ and the interaction free energy $F_{\rm int}$. We use
a standard lattice-gas formalism~\cite{andelmanjoanny} to derive the
demixing free energy of the surfactant monolayer:

\begin{eqnarray}
 \label{Fmeq}
  \frac{F_{\rm surf}}{k_BT}= b^{-2}\big[c\ln c
    +\left(1-c\right)\ln\left(1-c\right) \\
    \nonumber +\nu^{-1} c\left(1-c\right)
    -cZ\yy_s\big],
\end{eqnarray}
where $b^2$ is the close packing area of the surfactant head groups.
The first two terms account for the surfactant entropy of mixing,
where $c$ is the surfactant area fraction. The third term represents
a short-range interaction between the surfactants with $\nu^{-1}$ as
the interaction parameter. The fourth term is electrostatic
contribution of the charged surfactant layer, where $Z$ is the
valency of the surfactant head group (assumed henceforth to be
negative) and $\yy_s=e\psi(0)/k_BT$ is the surface potential on the
monolayer plane in units of $k_BT$.

%%%%%%%%%%%%%%%%%%%%%%%%%%%%%%%%%%%%%%%%%%%%%%%%%%%%%%%%%%%%%%%%%
% FIG 1 schematic setup

\begin{figure}[!]
  \includegraphics[width=70mm,clip,trim= 0 0 0 0]{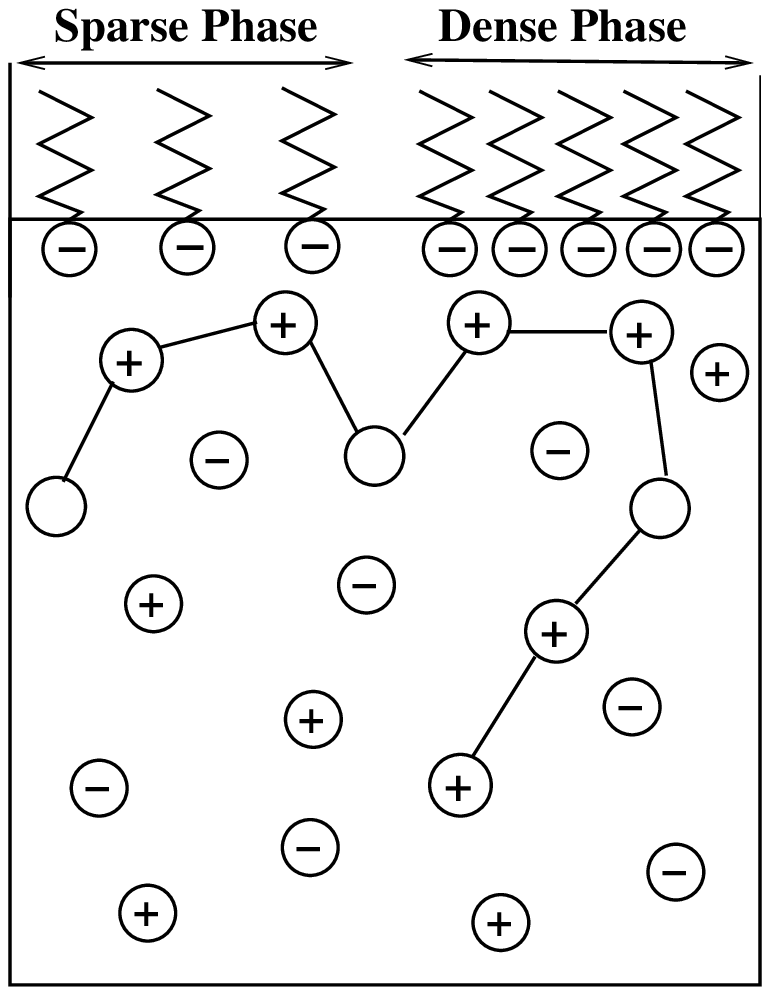}
  \caption{\footnotesize\textsf{A schematic view of the
    polyelectrolyte-surfactant system. The
    polyelectrolyte solution contains polyelectrolyte chains and
    monovalent salt (co- and counterions). The negatively charged
    surfactants are present only on the air-water interface, and
    phase separate into dense and dilute lateral phases.
  }}
\end{figure}

Without the fourth term, $F_{\rm surf}$ is the free energy commonly
used for neutral Langmuir monolayers. For simplicity, we take it to
be symmetric under the exchange of $c$ and $1-c$. The non-convexity
of $F_{\rm surf}$ implies that there is a demixing region with two
phases: (i) a dense phase with high surfactant concentration, and
(ii) a dilute phase with low surfactant concentration. The phase
demixing curve is determined by:

\begin{equation}
  \nu_b=\frac{1-2c}{\ln\left(1-c\right)-\ln c},
 \label{bino0}
\end{equation}
where $\nu=\nu_b(c)$  is the binodal line. Below a critical value
$\nu_c=1/2$ and for $\nu(c)<\nu_b(c)$  the monolayer phase separates
into two coexisting phases. Inside the coexistence region, another
line of interest is the spinodal line, satisfying $F_{\rm
surf}''(c)=0$. The spinodal line, separating the unstable and
metastable  regions with respect to surfactant density fluctuations,
is given by:

\begin{equation}
  \nu_{\rm sp}=2c\left(1-c\right).
 \label{surfspino}
\end{equation}
Its maximum coincides with that of the binodal line at the critical
point, $c_c=0.5$ and $\nu_c=0.5$. Therefore, examination of the
spinodal line gives a good initial approximation to the phase behavior.

The free energy of the bulk PE and salt solution, $F_{\rm
sol}=F_{\rm pol}+F_{\rm ions}+F_{\rm el}$, is more complex, and is
composed of three parts: (i) the polymer free energy $F_{\rm pol}$;
(ii) the small ion entropy $F_{\rm ions}$; and (iii) the
electrostatic free energy $F_{\rm el}$. The polymer free energy
is~\cite{itamar1}:

\begin{equation}
  \frac{F_{\rm pol}}{k_BT}=\int^\infty_0\dv x
  \left[\frac{a^2\phi_b^2}{6}\left|\frac{\dv \eta}{\dv x}\right|^2+
  \half{ v}_{\rm ex}\phi_b^4\left(\eta^2-1\right)^2\right],
 \label{Fpol}
\end{equation}
where the first term accounts for the polymer chain flexibility with
$a$ the monomer size, $\phi_b^2$ the bulk concentration of monomers
and $\eta^2(x)\equiv\phi^2/\phi_b^2$ the local monomer concentration
renormalized by its bulk value. The parameter $v_{\rm ex}$ is the
polymer excluded volume (second virial) coefficient. The second term
originates from the excluded volume interactions between the
monomers as well as the equilibrium condition of the PE solution in
contact with a bulk reservoir.

The change in small ion entropy with respect to the reservoir is:

\begin{equation}
  \frac{F_{\rm ions}}{k_BT}= \sum_{i=\pm}\int^\infty_0\dv x
    \left[ c_i\ln \frac{c_i(x)}{c_b^i}-c_i(x)+c^i_b\right]
 \label{Fion}
\end{equation}
where $c_{\pm}(x)$ is the local concentration of the positive and negative
ions and $c^{\pm}_b$ is their bulk concentration.

The last term of $F_{\rm sol}$, the electrostatic energy, is:

\begin{eqnarray}
 \label{Fel}
  \frac{F_{\rm el}}{k_BT} = 
	\qquad\qquad\qquad\qquad\qquad\qquad\qquad\qquad\qquad
	\\ \nonumber \qquad \int^\infty_0 \dv x
  \left[\left(c_+-c_-+f\phi_b^2\eta^2\right)\yy - \frac{1}{8\pi
  l_B}\left(\frac{\dv \yy}{\dv x}\right)^2\right]
\end{eqnarray}
where the dimensionless electrostatic potential (rescaled by the
temperature) is $\yy\equiv e\psi/k_BT$. The first three terms are
the interaction of the electrostatic potential with the positive
ion, negative ion and polyelectrolyte concentrations, respectively,
where the parameter $f$ is the fraction of charged monomers on the
PE chain. The last term is the self-energy of the electric field,
where $l_B=e^2/\varepsilon k_BT$ is the Bjerrum length. For water
with dielectric constant $\varepsilon=80$ at room temperature
$T=300$K, $l_B$ is equal to about $7$\,\AA.

The last contribution to the free energy, the non-electrostatic
interaction free energy $F_{\rm int}$, can be written as~\cite{chatellier}:

\begin{equation}
  \frac{F_{\rm int}}{k_BT}=-\half b^{-2}\eps\left(c-c^*\right)\phi_b^2\eta_s^2
 \label{Fint}
\end{equation}
In the above $\eta_s\equiv \eta(0)$ is the renormalized monomer
concentration at the surface, $x=0$. The strength of the
PE-surfactant interaction is defined by a phenomenological parameter
$\eps$, which has usints of volume. For positive values of $\eps$, 
the short-range interaction
of the PEs with the  surfactant is more favorable than those between
the PE and the bare interface, while for $\eps<0$ the opposite
occurs.  At concentration $c=c^*$ the surface is indifferent
(``special transition" line) ~\cite{deGennes,chatellier}. For
$\eps>0$ and low surfactant concentrations $c<c^*$, the short-range
interactions between the surface and the PE are repulsive, while for
higher surfactant concentrations $c>c^*$ these interactions are
attractive.

In order to find the equilibrium state of the PE-surfactant system
we minimize the total free energy, $F=F_{\rm surf}+F_{\rm
pol}+F_{\rm ions}+F_{\rm el}+F_{\rm int}$ with respect to the local
small ion densities $c_{\pm}$, electrostatic potential $\yy$ and
monomer order parameter $\eta$. This minimization yields the
following equations~\cite{itamar1,itamar2,us1}:

\begin{equation}
  \frac{\dv^2\yy}{\dv x^2}=\kappa^2\sinh \yy +k_m^2 \left(\e^\yy-
    \eta^2 \right)+\frac{4\pi l_B cZ}{b^2} \delta(x)\\
 \label{normPBmod}
\end{equation}
\begin{equation}
  \frac{a^2}{6}\frac{\dv ^2\eta}{\dv x^2}=\iv
    \left(\eta^3-\eta \right)+f \yy \eta -
    \frac{\eps\left(c-c^*\right)}{2b^2}\eta\delta(x)~,
 \label{normEdwardsmod}
\end{equation}
where  $\kappa^{-1}=\left(8\pi l_B \cs\right)^{-1/2}$  is the
Debye-H\"{u}ckel length scale for the screening of the electrostatic
potential in the presence of added salt and $k_m^{-1}=\left(4\pi l_B \phi_b^2
f\right)^{-1/2}$ is the corresponding length for the potential
decay due to counterions. Note that the actual decay of the electrostatic
potential is determined by a combination of salt, counterions,
and polymer screening effects.

The solution of Eqs.~(\ref{normPBmod}) and (\ref{normEdwardsmod})
requires four boundary conditions. Two of them are chosen at the
bulk, far from the surfactant monolayer, and two at the surfactant
layer. At the bulk we set the electrostatic potential as
$\yy(x\to\infty)=0$, while the monomer concentration is set equal to
the bulk monomer concentration $\eta(x\to\infty)=1$. The other two
boundary conditions are obtained by integrating
Eqs.~(\ref{normPBmod}) and (\ref{normEdwardsmod}) from $0$ to a
small distance  near the surface incorporating the surface
interactions of Eqs.~(\ref{normPBmod}),~(\ref{normEdwardsmod}):

\begin{eqnarray}
  \left.\frac{\dv \yy}{\dv x}\right|_{x=0} & = &\frac{4\pi l_BcZ}{b^2}
 \label{ybc} \\
  \left.\frac{\dv \eta}{\dv
  x}\right|_{x=0} & = & -\frac{3\eps\left(c-c^*\right)}{b^2a^2}\eta(0)
 \label{etabc}
\end{eqnarray}
Equation~(\ref{ybc}) is the usual electrostatic boundary
condition for a given surface charge density $\scharge\equiv
cZ/b^2$, while Eq.~(\ref{etabc}) is the Cahn - de Gennes boundary
condition~\cite{degennes}, which is often used for neutral
polymers~\cite{joanny}.

Equations~(\ref{normPBmod})-(\ref{etabc}) are solved numerically
using the relaxation method~\cite{nr}, as was described previously~\cite{us1}. 
For each surfactant concentration, we
calculate the value of  $\nu$ for which $F''(c)=0$, corresponding to
the spinodal curve. Then the full binodal line is calculated
numerically. We find that despite the electrostatic interactions
between the surfactants and the PE chains, the main characteristics
of the spinodal and binodal lines depend on the {\it short-range}
(non-electrostatic) PE-surfactant interactions. There are two main
regimes for the demixing curve. For large and positive $\eps(c-c^*)$
values (short-range attractive surface), the demixing curve is shown
to be always higher than the pure monolayer one,
Eq.~(\ref{surfspino}). For large negative $\eps(c-c^*)$ values
(short-range repulsive surface), the curve is always lower than the
pure one, Eq.~(\ref{surfspino}).

In order to incorporate the temperature dependence of the various
parts, we note that $\nu$, the inverse Flory-Huggins parameter, is
proportional to the temperature $T$. The chemical interaction
strength parameter $\eps$ is proportional to $T^{-1}$. The other
parts of the free energy have a weaker dependence on $T$, which can
be ignored in this simple model~\cite{chatellier}. Their
contribution is neither purely enthalpic nor purely entropic, and is
less significant. Therefore, in order to simulate the change in
temperature in our simple model, we follow the method shown
previously in Ref.~\cite{chatellier} and choose
$\nu\cdot\eps=\Theta$ to be constant.

%%%%%%%%%%%%%%%%%%%%%%%%%%%%%%%%%%%%%%%%%%%%%%%%%%%%%%%%%%%%%%%%%%%%%%%%%%%%%%
%%%%%%%%%%%%%%%%%%%%%%%%%%%%%%%%%%%%%%%%%%%%%%%%%%%%%%%%%%%%%%%%%%%%%%%%%%%%%%
\section{Numerical Results}
\label{phasedi}
%%%%%%%%%%%%%%%%%%%%%%%%%%%%%%%%%%%%%%%%%%%%%%%%%%%%%%%%%%%%%%%%%%%%%%%%%%%%%%
%%%%%%%%%%%%%%%%%%%%%%%%%%%%%%%%%%%%%%%%%%%%%%%%%%%%%%%%%%%%%%%%%%%%%%%%%%%%%%

%%%%%%%%%%%%%%%%%%%%%%%%%%%%%%%%%%%%%%%%%%%%%%%%%
\subsection{The Three PE-Surfactant Regimes}
%%%%%%%%%%%%%%%%%%%%%%%%%%%%%%%%%%%%%%%%%%%%%%%%%%

We find three distinct regimes for the PE-surfactant system. These
 regimes are presented in Fig~2, and discussed more
thoroughly in Sec.~\ref{analitical}. For repulsive surfaces, very
low $c$ and high $\cs$, the PEs deplete from the interface. The high
amount of salt screens the electrostatic interactions between the
constituents of the solution, and the free energy can be written as
a sum of three terms: an interface term, a neutral polymer term and
a salt solution term. For higher $c$, short-range repulsive surfaces
and low salt, the PEs adsorb electrostatically on the surface. In
this case, the free energy is dominated by the electrostatic
attraction between the PE chains and the surfactant layer. For the
opposite case of sort-range attractive surfaces, the adsorption is
dominated by the short-range attraction.

In Fig.~2a we present the phase diagram in the plane ($\cs$, $c$)
differentiating the three regimes in the case of $\eps>0$, namely,
when the surfactant-PE interaction is more favorable than the PE -
bare surface interaction. For $c<c^*$, in the area marked E, the
short-range interaction between the surface and the PE chains is
repulsive, and the adsorption is dominated by the electrostatic
attraction between the PEs and the surfactants. For high amounts of
added salt, this electrostatic attraction is screened, and the PE
chains deplete (marked D). For $c>c^*$, (marked C) the PE chains
chemically adsorb on the surface for all $\cs$. The electrostatic
interactions in this region are dominated by the monomer-monomer
electrostatic repulsion, and hence the addition of salt increases
the adsorption, rather than causing depletion. In Fig.~2b we show
the opposite case of $\eps<0$. In this case the line separating the
phases is no longer monotonic. For $c<c^*$, the adsorption is mainly
chemical, and is enhanced by addition of salt. For $c>c^*$, the
adsorption becomes electrostatic and depends on $\cs$. The amount of
salt necessary for depletion decreases with $c$ for low $c$ since
the increase in $c$ increases the short-range repulsion between the
surface and the PE chains, while the surface charge is not strong
enough to electrostatically attract the PEs. For higher $c$, the
increase in the surface charge with $c$ increases mainly the
electrostatic adsorption, and the necessary amount of salt increases
with $c$.

%%%%%%%%%%%%%
%% Fig 2

\begin{figure}[!htp]
\includegraphics[keepaspectratio=true,width=110mm,clip=true, trim=200 0 100 0]{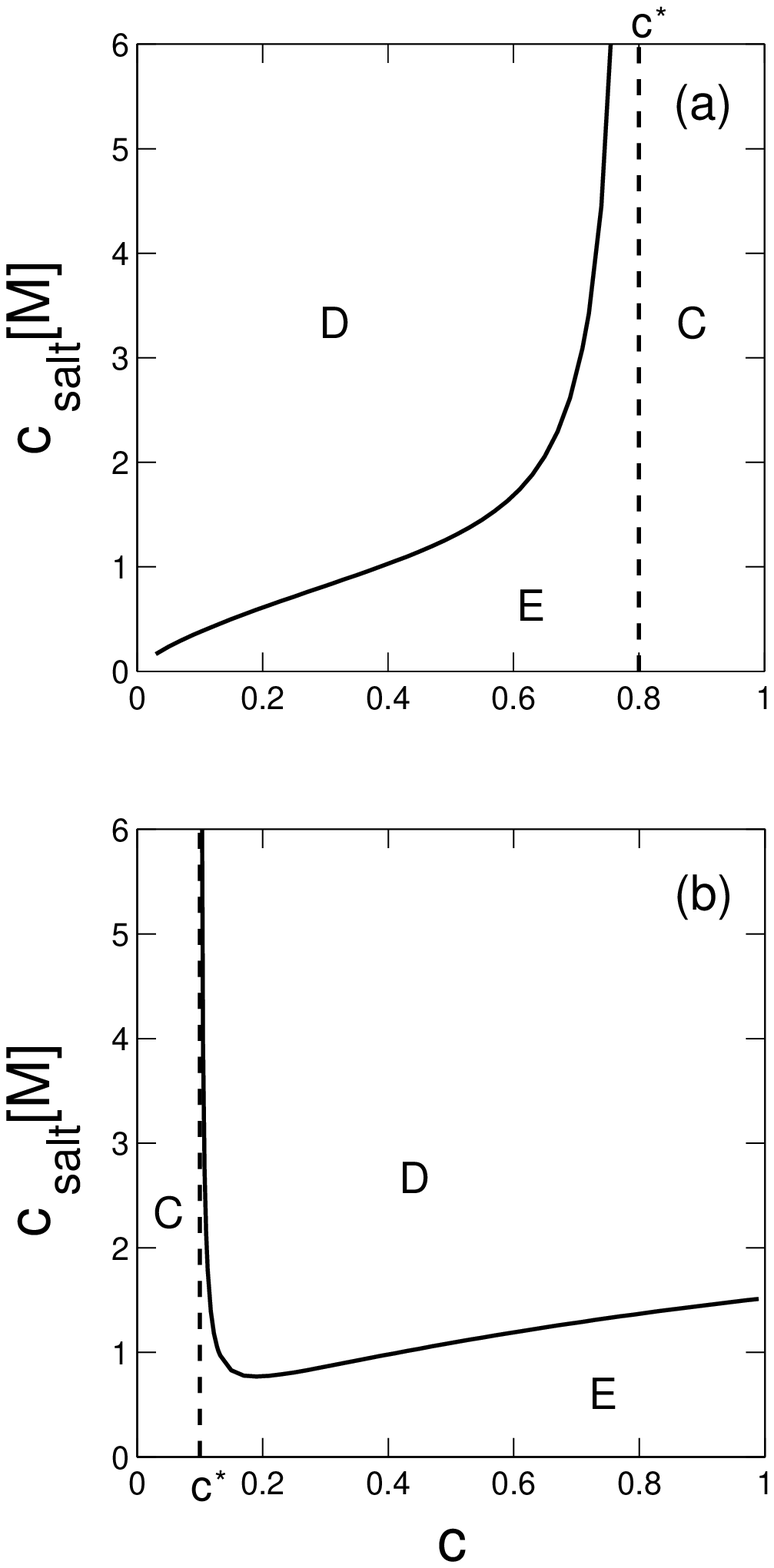}

  \caption{\footnotesize\textsf{ The numerically calculated
     phase diagram of the three PE-surfacant regimes
     are presented for two values of $\nu\cdot\eps$ and $c^*$. In both
     figure parts
     the solid line differentiates the electrostatic adsorption
     regime (marked E)
     from the depletion regime (D), and the dashed line differentiates both
     of them from the chemical adsorption regime (marked C). In (a) we set
     $\nu\cdot\eps=0.5b^2a$ and $c^*=0.8$, meaning that the short-range
     interaction of the PE chains with the charged surfactants is more
     favorable. For $c<c^*$ and
     low salt concentrations, the electrostatic attraction between the
     surfactants and  the PE chains causes them to adsorb on the
     monolayer. For higher salt concentrations the PE and surface charges
     are screened by the salt ions, and the PE chains deplete. For $c>c^*$,
     the short-range attraction between the surfactants and the PE chains
     is strong enough to allow PE adsorption for any salt concentration.
     In (b) we show the opposite case of $\nu\cdot\eps=-0.9b^2a$ and
     $c^*=0.1$, where the short-range interactions with the surfactants are
     less favorable. Here for $c<c^*$ the short-range interactions
     cause the PE to adsorb, while for $c>c^*$ the PE chains adsorb or
     deplete according to the strength of the electrostatic interactions
     and the added salt concentration. In both plots we use $\nu=0.1$,
     $a=5$\,\AA, $v_{\rm ex}=10$\,\AA$^3$, $\phi_b^2=10^{-8}$\,\AA$^{-3}$,
     $Z=1,\, f=0.5$,
     $b=10\,$\,\AA, $T=300$\,K and $\varepsilon=80$.
}}

\end{figure}

%%%%%%%%%%%%%%%%%%%%%%%%%%%%%%%%%%%%%%%%%%%%%%%%%%
\subsection{The Demixing Curve}
%%%%%%%%%%%%%%%%%%%%%%%%%%%%%%%%%%%%%%%%%%%%%%%%%%

{\em Repulsive Surface}~~~The spinodal lines of repulsive
PE-surfaces ($c^*=3.0,\,\nu\cdot\eps=0.5b^2a$) for several values of
$\cs$ are presented in Fig.~3a. The spinodal of the pure surface
monolayer, Eq.~(\ref{surfspino}), is shown to be always higher than
any of the charged-surface spinodals. All three spinodal lines are
almost symmetric under the change $c\rightarrow 1-c$, and the critical 
points are very close to the
\newpage

\begin{widetext}

%%%%%%%%%%%%%
%% FIG 3
\begin{figure}[!ht]
\includegraphics[keepaspectratio=true,width=183mm,clip=true,trim=50 120 0 150]{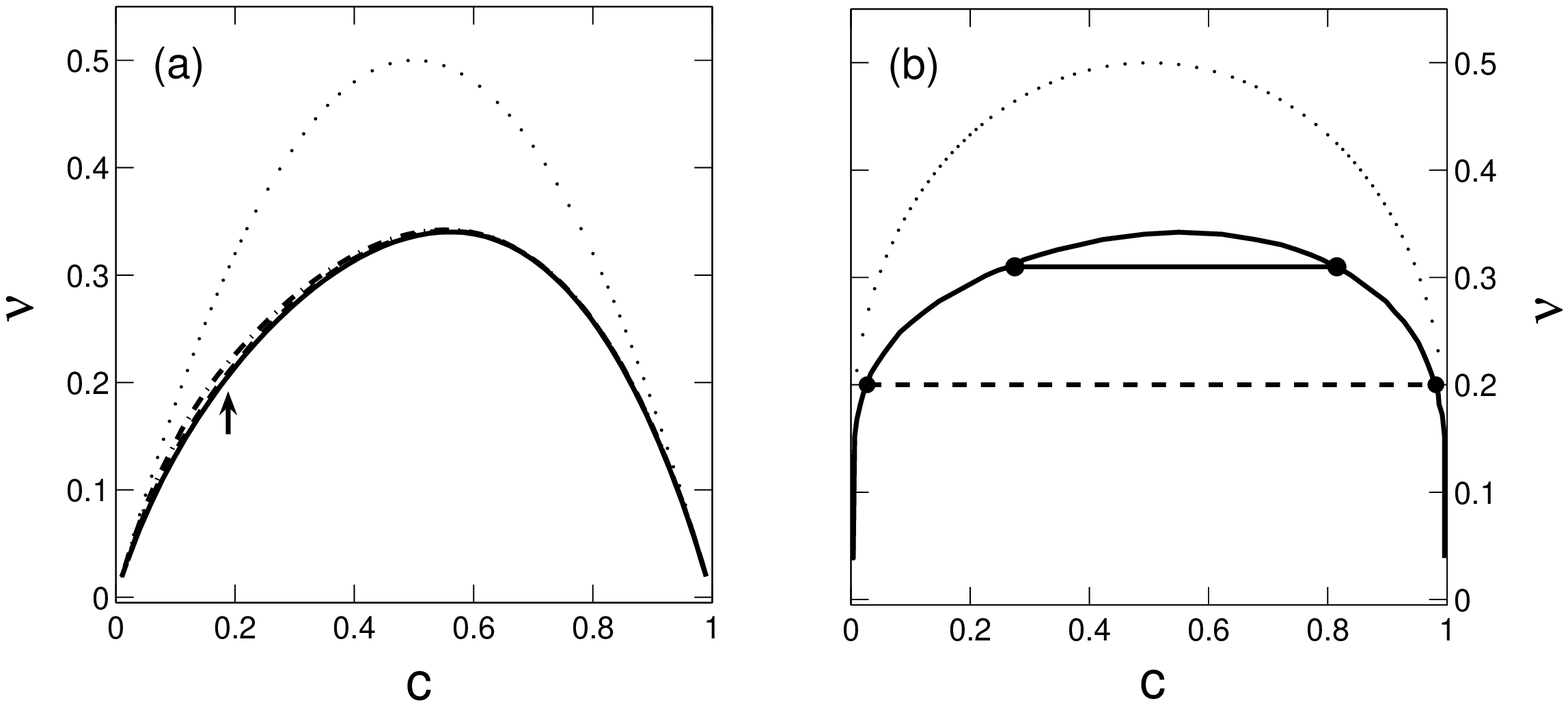}

  \caption{\footnotesize\textsf{
    (a) The spinodal lines of a repulsive surface system are
    presented as function of the surface coverage $c$. The spinodal is
    shown for three amounts of added salt. The solid line corresponds to
    $\cs=0.1$\,M, the dashed line to $\cs=0.5$\,M and the dashed-dotted line to
    $\cs=1.0$\,M. All lines share $c^*=3,\,a=5\,$\,\AA,
    $v_{\rm ex}=10$\,\AA$^3$, $\phi_b^2=10^{-8}$\,\AA$^{-3}$, $Z=1,\, f=1$,
    $b=10$\,\AA,
    $\nu\eps=0.5b^2a$, $T=300$\,K, $\varepsilon=80$. All three spinodal lines
    are seen to be almost identical, differing only in the region close
    to the adsorption-depletion crossover, marked by an arrow.
    The dotted line marks the pure monolayer spinodal, Eq.~(\ref{surfspino}),
    and is shown for comparison.
    (b) The binodal (phase separation) line of a repulsive surface
    system is presented as function of the surface coverage $c$ for
    $\cs=1.0$\,M. Other parameters are the same as in (a).
    The solid line is the binodal, while the dashed and dashed-dotted
    lines connect the separated phases for $\nu=0.2$ and $\nu=0.31$,
    respectively. The dotted line corresponds to the pure-monolayer
    binodal, Eq.~(\ref{bino0}), and is shown for comparison. }}

\end{figure}

\end{widetext}
pure surface one at $c_c=0.5$.
In Fig.~3b we present the binodal line of a repulsive surface for
$\cs=1.0$\,M. The binodal line of the pure surface,
Eq.~(\ref{bino0}), is also presented, and is higher than the
numerical binodal for all $c$. Like the spinodal line, the numerical
binodal is almost symmetric with respect to the change $c\rightarrow
1-c$.
For high $c$ values, all spinodal lines of Fig.~3a overlap almost
identically, showing that the dependence of the spinodal on the
amount of added salt is very weak. For low $c$ values, careful
observation of the three spinodal lines shows a difference between
the lines, where the low salt lines are higher than the high salt
ones. The three lines join together at $c\sim0.22$ (see arrow). This
difference between the spinodal lines results from the PE depletion
at low $c$. For low $c$ values, the spinodal resembles a depletion
spinodal, where addition of salt has a strong effect on the
spinodal. For higher $c$ values, the surface charge is high enough
to allow PE adsorption. All three lines then move toward the
electrostatic adsorption spinodal, that is weakly dependent on salt.
The point where all lines join together is where the high salt
spinodal crosses over from the depletion regime to the electrostatic
adsorption regime, and its numerical value agrees with the
adsorption-depletion crossover at $1.0$\,M of salt taken from
Fig.~4. The spinodal lines in the different regimes are discussed
further in Sec.~\ref{analitical}.

In Fig.~4 the adsorbed amount $\Gamma\equiv\phi_b^2\int_0^\infty\dv
x(\eta^2-1)$ is plotted as a function of the surfactant surface
coverage $c$ for $\cs=1.0$\,M. We set $\nu=0.2,\,0.31$ in Figs.~4a,
4b, respectively. The adsorption is seen to be positive only from a
threshold surface coverage $c_{\rm dp}$, which is similar to the
crossover point between spinodal lines shown in Fig.~3. Above this
threshold, the adsorbed amount increases linearly with the surface
coverage, in agreement with previous results~\cite{us1,us2}. The
threshold is found not to change with $\nu$.

It is interesting to note the effect of the adsorption-depletion
crossover on $\Gamma$ in the coexistence region. In Fig.~3b we mark
on the binodal line the coexisting phases for $\nu=0.2$ (dashed
line) and for $\nu=0.31$ (dashed-dotted line).  In Fig.~4 we show
$\Gamma$ for the same coexisting phases. Using the lever rule
construction~\cite{chatellier},  $\Gamma$ inside the phase
coexistence lies on a straight line between the two marked phases.
For $\nu=0.2$ (dashed line in Fig.~4a), the PE adsorbs in one of the
phases while depleting from the other. As a result, the adsorption
in the  phase separated region  is always higher than onto a
reference single phase (solid line), having the same mean surface
coverage $c$,
For $\nu=0.31$ the PE adsorbs on the two coexisting phases. The
adsorption in the phase separated region (dashed-dotted line in
Fig.~4b) is almost identical to the adsorption to the single
corresponding mixed phase (solid line), and there is no gain in the
adsorption due to demixing.

%%%%%%%%%%%%%
%% FIG 4

\begin{figure}
\includegraphics[keepaspectratio=true,width=105mm,clip=true,trim=180 0 120 0]{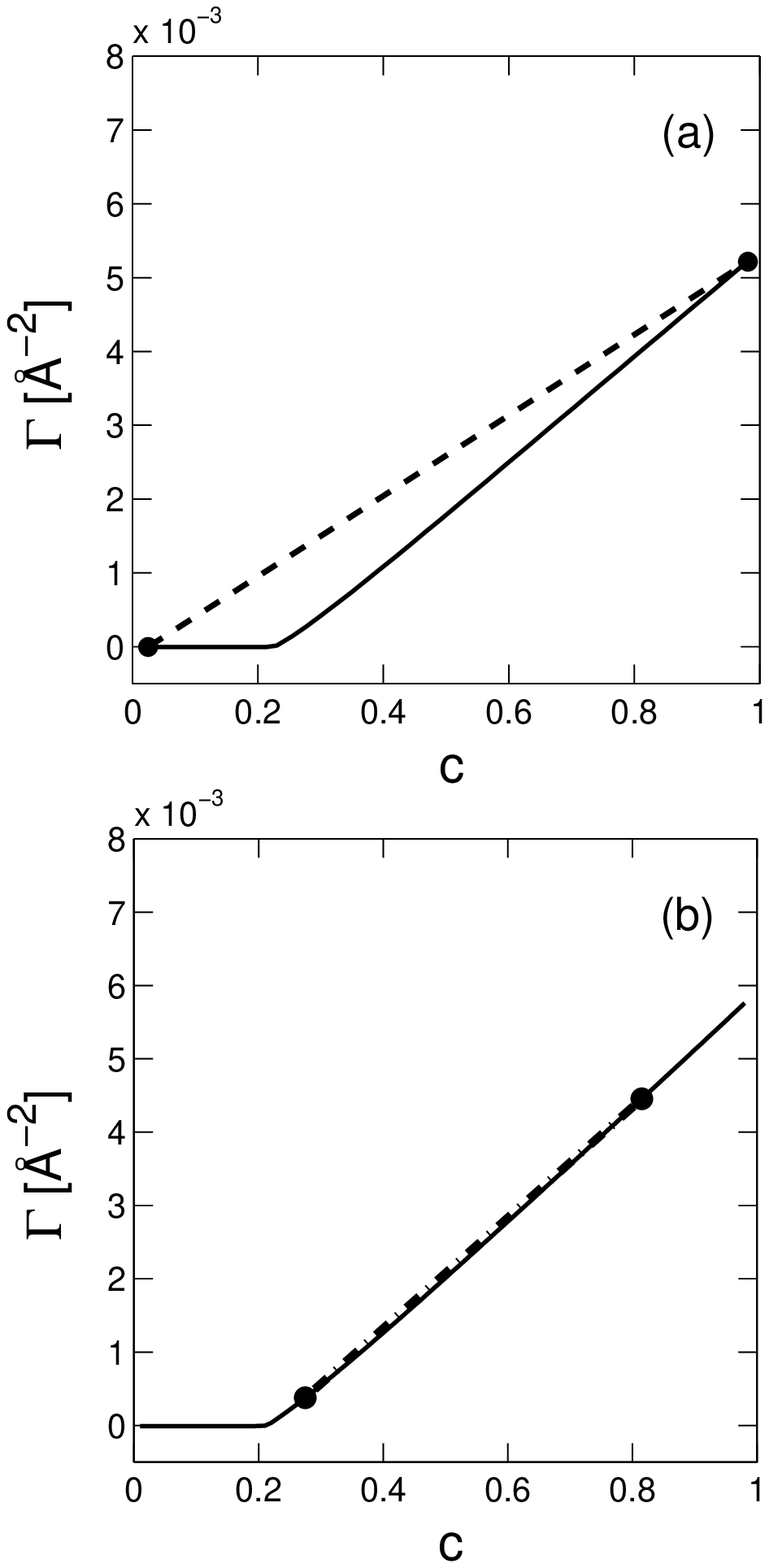}

  \caption{\footnotesize\textsf{
   The adsorbed amount $\Gamma$ is plotted as a function of the
    surfactant coverage for two  values of $\nu$.
    In (a) the solid line correspond to $\nu=0.2$ while in (b) $\nu=0.31$.
    Both figures share $\cs=1.0$\,M, and other parameters are the same as in
    Fig.~3. In both graphs, a low $c$ depletion regime can be found, in which
    $\Gamma$ is slightly negative. For larger surface coverage,
    the adsorbed amount scales linearly with the surface coverage, and no
    considerable convexity is obtained. The threshold surface coverage
    for adsorption is unaffected by $\nu$. The dashed
    line in (a) and dashed-dotted line in (b) are the adsorbed amounts
    inside
    the phase coexistence region, and are constructed by the lever rule.
    A significant increase in adsorption can only be achieved
    when one of the separated phases is a depletion phase. This is
    the case in (a). For higher values of $\nu$ (b), the phase separated
    adsorption line
    is almost the same as the original (single phase) adsorption line, and
    no gain in adsorption is achieved.
}}

\end{figure}

The difference between these two behaviors can be qualitatively
explained by the following argument. In the electrostatic adsorption
case, the adsorbed amount of monomers scales linearly with the
surface charge~\cite{itamar1}, and hence also with $c$. The
adsorption of the PE chains aims mainly to balance the surface
charge, and since the demixing does not change the overall surface
charge, the total adsorbed amount does not change as well. In the
case of a depletion phase- adsorption phase coexistence, the charges
on the low $c$ (depletion) domain are screened by salt ions, while
the higher surface charge domains are screened by the PE chains.  In
this case, even if the total surface charge is too weak to allow PE
adsorption in a single mixed phase, the phase separation allows
adsorption at a part of the surface area, and thus allows for a
positive adsorbed amount. The crossover between these two behaviors
is at $\nu\simeq0.3$, corresponding to the low $c$  phase being at
$c\simeq c_{\rm dp}$. Namely, that the low $c$ phase is at the
adsorption-depletion transition.

{\em Short-Range Attractive Surface}~~~The opposite case of a
short-range attractive surface is presented in Fig.~5. As can be
seen from the figure, the spinodal line is much higher than the
corresponding pure-surface spinodal line, and it is far less
symmetrical. The addition of salt is shown to increase the spinodal
temperature. This increase with salt can be explained by both the
screening of the electrostatic repulsion between the lipids, and by
the increase in the adsorbed amount of monomers in this
case~\cite{us2}, which increases the effective attraction between
the lipids. The corresponding adsorbed amount of monomers is shown
in Fig.~6. The adsorbed amount here is convex in $c$, showing that
the adsorption to a demixed two-phase system is always larger than
to a single phase system.

{\em Attractive-Repulsive Crossover}~~~In Fig.~7 we show the
spinodal lines of monomers for $c^*=0.7$ and both $\cs=1.0$\,M and
$\cs=0.1$\,M. It is easily seen that the low salt spinodal is higher
than the high salt one for $c<c^*$ (repulsive surface), while the
opposite  occurs for $c>c^*$, consistent with Fig.~5. Although all
three regimes occur in this spinodal, the curve has only one
critical point, located near the crossover between the spinodal
behavior of the electrostatic adsorption limit to the chemical
adsorption limit, at $c\simeq c^*$. This shows that the transition
of the spinodal line between the regimes is gradual rather than
abrupt. The $\cs=1.0$\,M line changes convexity at $c\sim 0.2$,
creating a ``shoulder" in the spinodal. This shoulder marks the
transition from a depletion regime to an electrostatic adsorption
regime. For extremely strong short-range interactions (very high
$\eps$), this shoulder may transform into a triple point on the
spinodal. However, we did not find such a triple point numerically
for reasonable values of our parameter.

%%%%%%%%%%%%%
%% FIG 5
\begin{figure}
\includegraphics[keepaspectratio=true,width=110mm,clip=true,trim=60 0 -80 0]{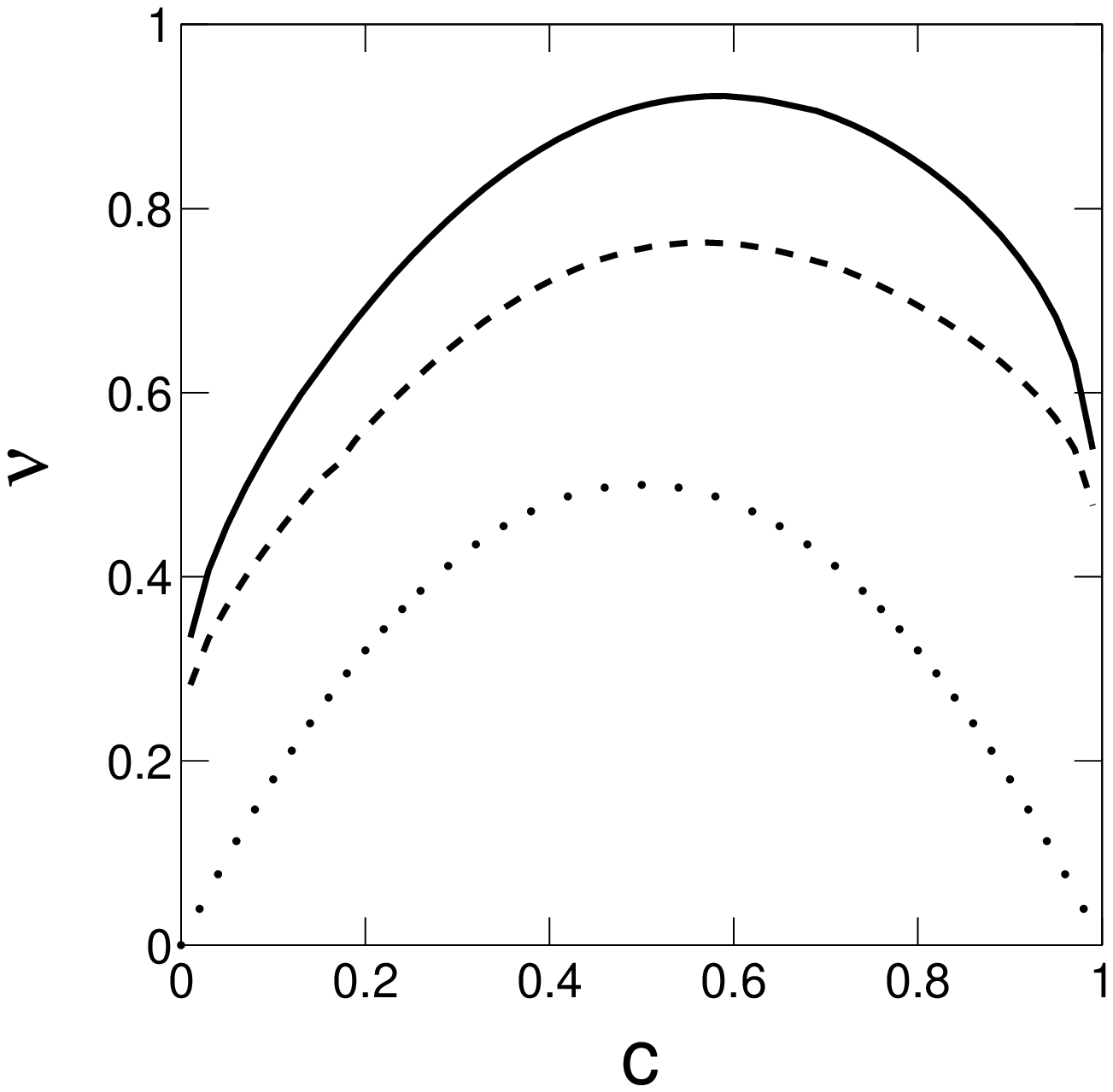}
  \caption{\footnotesize\textsf{
    The spinodal line for a short-range attractive surface, and two
    salinities: $\cs=1$\,M (solid line) and $\cs=0.1$\,M (dashed line).
    The dotted line corresponds to
    the pure surface spinodal, Eq.~(\ref{surfspino}). The profiles
    were calculated for $c^*=-1$ and $Z=0.1$, and other parameters
    are the same as in Fig.~3. In contrast to the repulsive surface
    case, here the PE adsorption pushes the spinodal to higher
    temperatures than the spinodal of the pure monolayer. The
    effect of salt is also more pronounced than in the repulsive surface
    case, and it is shown that addition of salt increases the spinodal
    temperature considerably.}}
\end{figure}

In Fig.~8 the adsorbed amount for $c^*=0.7$ is plotted against the
surface coverage $c$. For low $c$ values, we see the PEs deplete,
similar to Fig.~4, while for higher $c$ the adsorption crosses over
from a linear electrostatic adsorption profile to a convex chemical
adsorption profile. The addition of salt decreases the adsorption
for low $c$ values, and increases it for higher surface coverage.
Note, however, that the crossover between the two salt dependencies
is not at $c=c^*$ but for a lower $c$. This result is in agreement
with a previous one~\cite{us2}, which shows that when the
short-range repulsion between the surface and the PE chains becomes
small enough, the adsorbed amount starts to increase with $\cs$ even
before the surface becomes attractive, due to an increase in the
surface monomer concentration $\eta_s^2$.

%%%%%%%%%%%%
%% FIG 6
\begin{figure}
\includegraphics[keepaspectratio=true,width=108mm,clip=true,trim=40 0 -50 0]{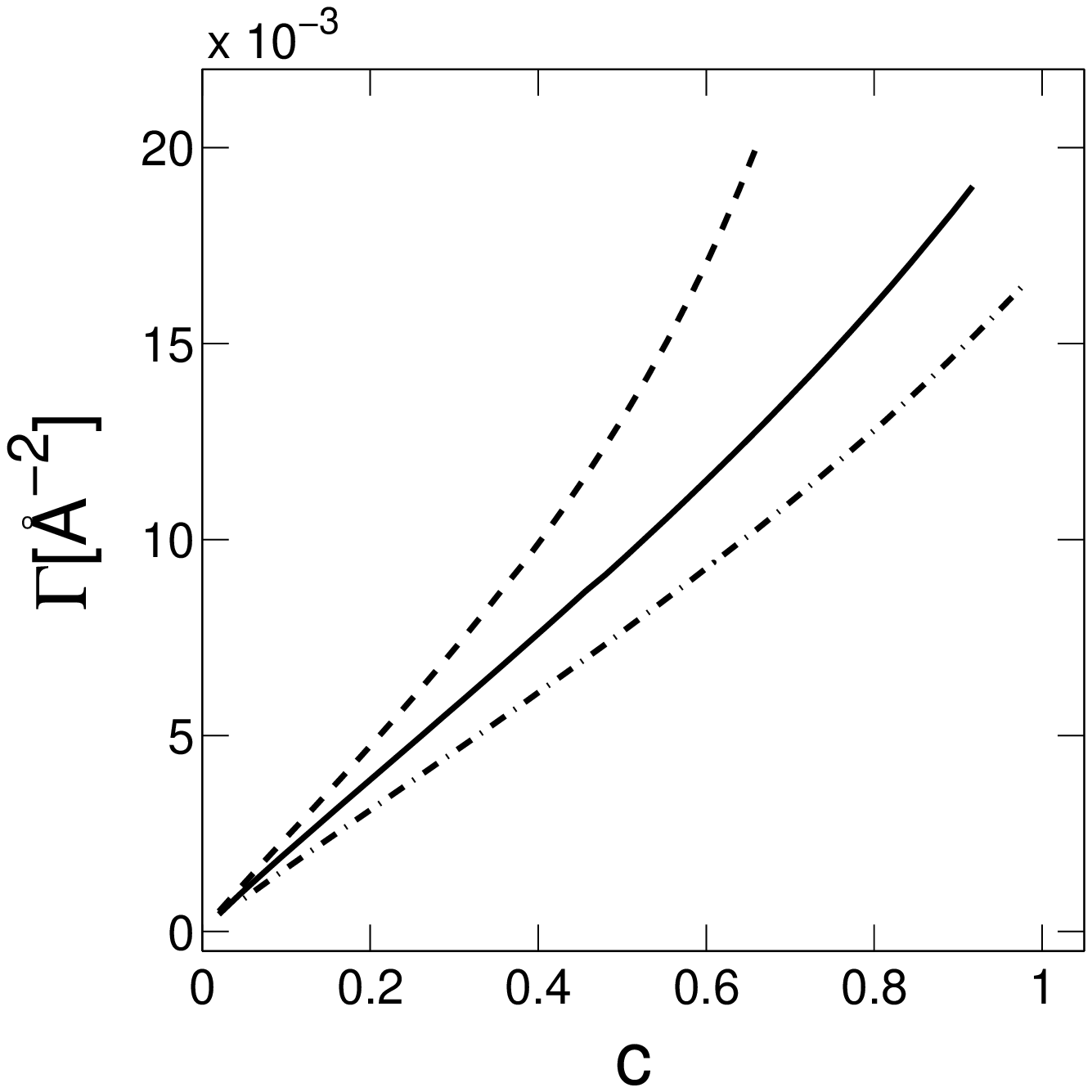}
  \caption{\footnotesize\textsf{
    The adsorbed amounts of monomers are shown for the
    short-range attractive surface. The dashed line corresponds to
    $\cs=1$\,M, $\nu=0.3$, the solid
    line to $\cs=0.5$\,M, $\nu=0.4$  and the dashed-dotted line to
    $\cs=0.5$\,M, $\nu=0.5$. All profiles share $c^*=0$, and other parameters
    are the same as in Figs.~3. In contrast to the repulsive surface
    case, here the addition of salt increases the adsorption. An increase
    in the temperature $\nu$ decreases the adsorption as expected. The
    adsorption profiles $\Gamma(c)$ are convex with respect to $c$, showing
    that a demixed lipid monolayer adsorbs more PEs than onto a homogeneous
    layer with the same mean coverage $c$.}}

\end{figure}

So far we discussed the case where the short-range interactions
between the PE chains and the surfactant are more favorable than
with the bare surface ($\eps>0$). In the case of a neutral
polymer, changing the sign of $\eps$ is exactly equivalent to
replacing $c^*\rightarrow 1-c^*$ and $c\rightarrow 1-c$. However,
in our case the fact that the surfactant and polymers are
oppositely charged breaks this symmetry.

For very strong short-range interactions (large negative
$\nu\cdot\eps$) and $0<c^*<0.5$ , a triple point can emerge in the
following way. For low $c$ values, the spinodal resembles the
attractive surface spinodal. The spinodal line in this region is
always higher than the pure surface spinodal Eq.~(\ref{surfspino}),
as was shown in Fig.~5. For $c_{\rm dp}>c>c^*$, the spinodal line
decreases substantially to the depletion spinodal, and for $c>c_{\rm
dp}$ it moves to the electrostatic adsorption spinodal. Both of the
latter spinodal lines are similar to those in Fig.~3a, both are
lower than the pure surface spinodal, and have a critical point at
$c_c\simeq 0.5$. Therefore, there must be a triple point at $c<0.5$,
connecting three coexisting phases: (i) a low $c$
chemical-adsorption phase; (ii) a low adsorption phase; and (iii) an
electrostatic adsorption phase, as shown in Fig.~9.

In Fig.~9 we show the numerically calculated triple point
corresponding to the above conditions. At high temperatures $\nu
>0.29$, the phase coexistence for low surface coverage connects a
chemical-adsorption phase with either another chemical adsorption
phase, or a depletion phase. For higher surface coverage, the phase
coexistence connects either a depletion phase or an electrostatic
adsorption phase with another electrostatic adsorption phase. A
triple point connecting all three phases occurs for $c_{\rm
tr}\simeq0.22,\, \nu_{\rm tr}=0.29$.

%%%%%%%%%%%%
%% FIG 7
\begin{figure}
\includegraphics[keepaspectratio=true,width=105mm,clip=true,trim=60 0 -50 0]{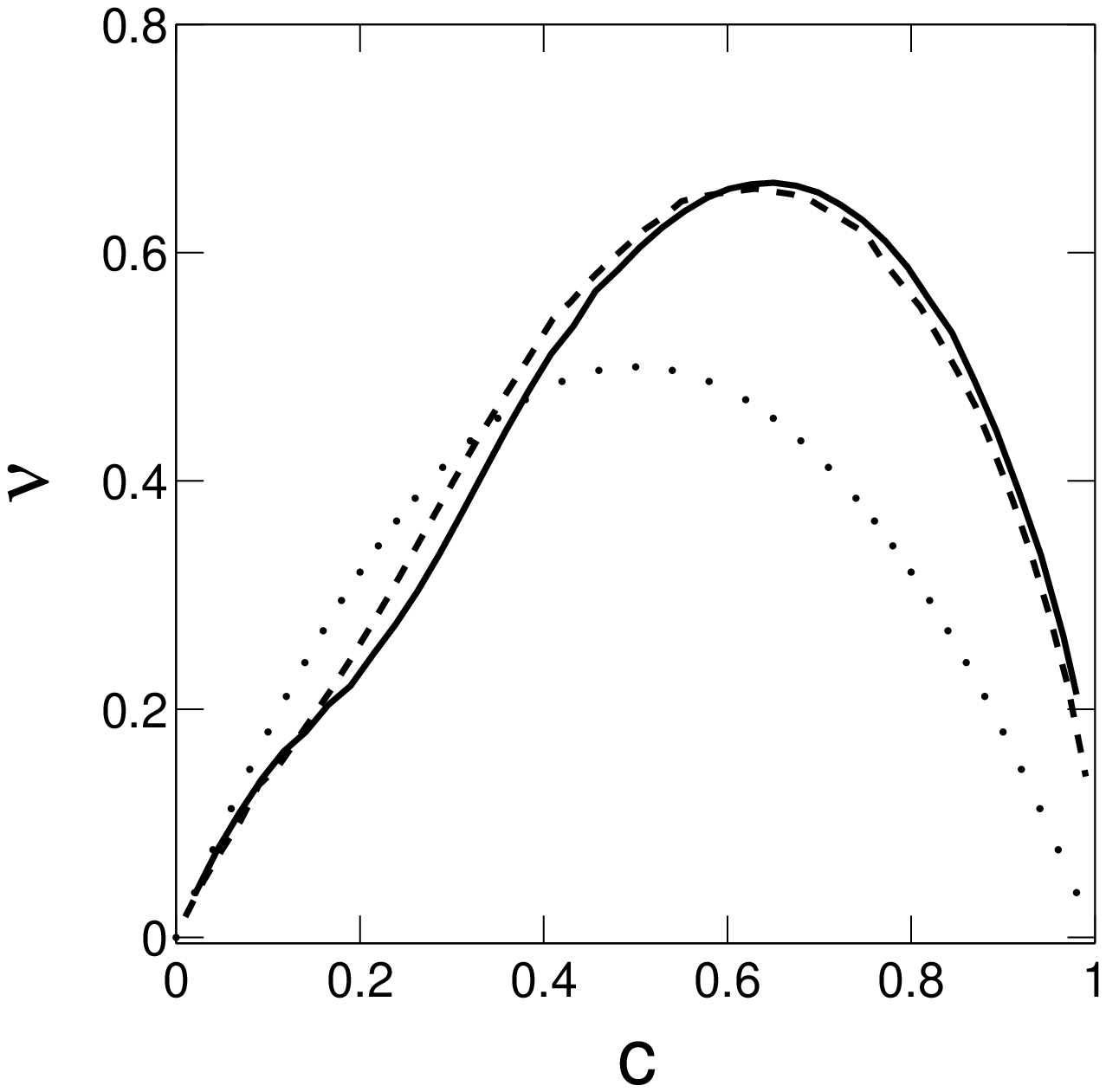}
  \caption{\footnotesize\textsf{
    The spinodal line is presented for $c^*=0.7$ for two salinities:
    $\cs=0.1$\,M (dashed line) and $\cs=1.0$\,M (solid line).
    All other parameters are the same as in Fig.~3. The dotted line is
    the pure surface spinodal, and is presented for comparison.
    Both spinodals are very close, but one can see that the low salt
    spinodal line is slightly higher than
    the high salt spinodal line for $c<c^*$, and slightly lower than it
    for $c>c^*$. The saddle point on the high salt spinodal line corresponds
    to the adsorption-depletion crossover, and occurs at the same surface
    coverage as in the repulsive surface case.
    Both spinodals cross over from a low $\nu$ range (lower than the
    pure surface spinodal) to a high $\nu$ range (resembling an
    attractive surface) at $c\sim 0.4$.}}

\end{figure}

%%%%%%%%%%%%
%% FIG 8
\begin{figure}
\includegraphics[keepaspectratio=true,width=105mm,clip=true,trim=50 0 -50 0]{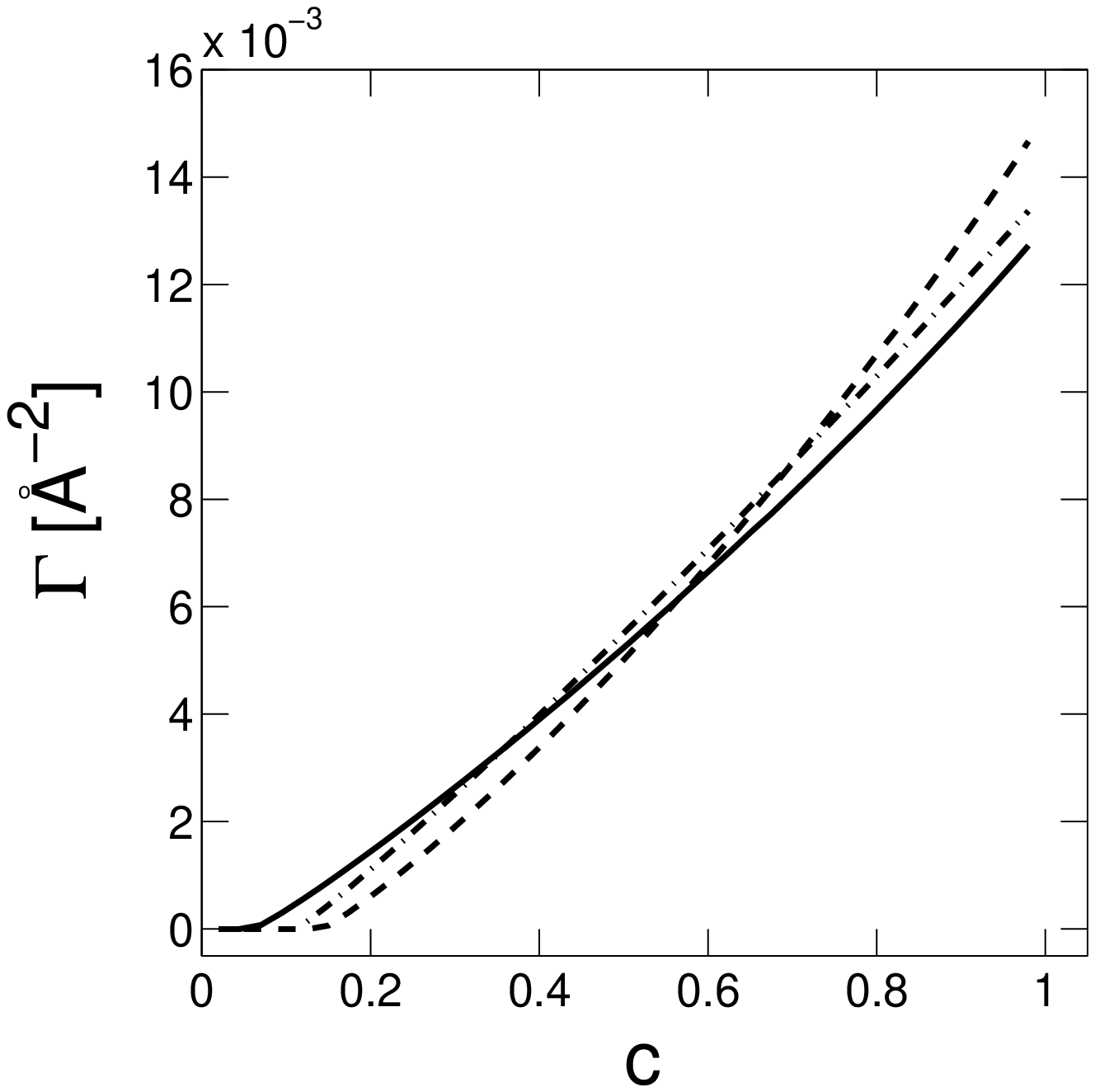}
  \caption{\footnotesize\textsf{The adsorbed amount of monomers $\Gamma$
    is shown
    for $c^*=0.7$. The solid line corresponds to
    $\cs=0.5$\,M, $\nu=0.3$, the dashed-dotted line to $\cs=1$\,M, $\nu=0.5$
    and the dashed line to $\cs=1$\,M, $\nu=0.3$. All other parameters
    are the same as in Fig.~7.
    The increase in the amount of salt causes the adsorbed amount to
    decrease for low $c$, and to increase for high $c$.
    }}

\end{figure}

%%%%%%%%%%%%%%%%%%%%%%%%%%%%%%%%%%%%%%%%%%%%%%%%%%%%%%%%%%%%%%%%%%%%%%%%%%%%%%%%%%
%%%%%%%%%%%%%%%%%%%%%%%%%%%%%%%%%%%%%%%%%%%%%%%%%%%%%%%%%%%%%%%%%%%%%%%%%%%%%%%%%%
\section{Analytical Results}
\label{analitical}
%%%%%%%%%%%%%%%%%%%%%%%%%%%%%%%%%%%%%%%%%%%%%%%%%%%%%%%%%%%%%%%%%%%%%%%%%%%%%%%%%%
%%%%%%%%%%%%%%%%%%%%%%%%%%%%%%%%%%%%%%%%%%%%%%%%%%%%%%%%%%%%%%%%%%%%%%%%%%%%%%%%%%

In this section we present a theory to explain the three
PE-surfactant regimes, and, in particular, aimed at explaining the
physics behind the spinodal behavior in each regime. We derive an
approximate form for the free energy in each regime, and show the
resulting spinodal line.  Our model is based on the scaling results
of Refs.~\cite{itamar1,us1}, and is shown to capture the correct
behavior of the spinodal lines in each limit. The analytical
equations can be compared to the numerical results with good
accuracy, as shown below.

%%%%%%%%%%%%%%%%%%%%%%%%%%%%%%%%%%%%%%%%%%%%%%%%%%%%%%%%%%%%%%%%%%%%%%%%%%%%%%%%%%
\subsection{Polyelectrolyte Depletion Regime}
\label{pdrepsdep}
%%%%%%%%%%%%%%%%%%%%%%%%%%%%%%%%%%%%%%%%%%%%%%%%%%%%%%%%%%%%%%%%%%%%%%%%%%%%%%%%%%

When $\eps\left(c-c^*\right)<0$ the surface repels the PE chains,
and thus the only driving force for adsorption is the electrostatic
attraction between the surfactants and the PE chains. This
electrostatic attraction strongly depends on the amount of small
ions in the solution~\cite{us1}. When this amount increases beyond a
threshold value, the salt ions neutralize the surface and PE
charges, and the solution behaves like a neutral polymer solution.
In this case, the repulsive short-range interaction between the PE
chains and the surface causes the PE chains to desorb entirely from
the surface. This threshold value of salt was previously
shown~\cite{us1} to depend on the surface charge density of the
interface and on the charge of the PEs via the scaling law
$(\cs^*)^{3/2}\sim \scharge f l_B^{-1/2}a^{-2}$. This relation can
be inverted to give a threshold surfactant concentration $c_{\rm
dp}$:

\begin{equation}
  c_{\rm dp} \simeq A_{\rm dp}\frac{\cs^{3/2}l_B^{1/2}a^{2}b^2}{Z f}
 \label{cdp}
\end{equation}
below which the PE chains do not adsorb on the charged surface. The
prefactor $A_{\rm dp}$ can be calculated by comparison of
Eq.~(\ref{cdp}) with the numerical results as in Figs.~3-4, yielding
$A_{\rm dp}\simeq 2.35$.

When the PE chains desorb from the surface, the electrostatic
potential depends only on the adsorption profile of the small ions.
According to the  well known Poisson-Boltzmann
theory~\cite{andelmanmembranes}, this decay is exponential as
function of the distance from the surface $x$. This exponential
decay causes the interaction term  $f\phi_b^2\eta^2\yy$ between the
PE chains and the electrostatic potential in Eq.~(\ref{Fel}) to be
small in comparison to the other terms. The free energy hence
decouples and describes a neutral polymer and an ionic solution.
Using the Debye-H\"{u}ckel approximation for the small electrostatic
potentials, the total free energy becomes:

\begin{eqnarray}
 \label{Fsmallion}
  F & = &F'_{\rm surf}+\frac{2\pi l_B c^2Z^2}{\kappa b^4}- \\ \nonumber
  & & b^2 \frac{a^2v_{\rm ex}\phi_b^4}
    {18\eps\left(c^*-c\right)}
  \left(3+\frac{b^4a^2v_{\rm ex}\phi_b^2}{3\eps^2\left(c^*-c\right)^2}\right)
\end{eqnarray}
The second term is the free energy of a solution of small ions
under the Debye-H\"{u}ckel approximation, the third term is the
free energy of a neutral polymer solution as in
Ref.~\cite{chatellier}, and the first term $F'_{\rm surf}$ is the
free energy of the neutral surfactant monolayer:

\begin{eqnarray}
  F'_{\rm surf}=b^{-2}\big[c\ln c+\left(1-c\right)\ln\left(1-c\right)
     \\ \nonumber +
    \nu^{-1}c\left(1-c\right)\big]
 \label{Fsurf}
\end{eqnarray}
where we note that the surface charge term was already taken into account in
the small ion solution term of Eq.~(\ref{Fsmallion}).
Differentiating this free energy twice
with respect to $c$, we obtain the following equation for the spinodal:

\begin{eqnarray}
 \label{SpinDep}
  b^{-2} \left(\frac{1}{c}+\frac{1}{1-c}-2\nu_{\rm sp}^{-1}\right) +
    \frac{4\pi l_B Z^2}{\kappa b^4}\qquad & &\\ \nonumber
    - b^2 \frac{a^2v_{\rm ex}\phi_b^4}{18\eps\left(c^*-c\right)^3}
  \left(6+\frac{4b^4a^2v_{\rm ex}\phi_b^2}
  {\eps^2\left(c^*-c\right)^2}\right)&=&0
\end{eqnarray}
The first term of the spinodal Eq.~(\ref{SpinDep}) is the pure
surfactant monolayer spinodal as in Eq.~(\ref{surfspino}), while
the second term accounts for
the electrostatic interactions between the salt ions and the
surface and the third term for the neutral polymer
solution~\cite{chatellier}. The  neutral polymer term, as was shown
in Ref.~\cite{chatellier}, causes the spinodal line to
increase slightly. The magnitude of this term is inversely proportional to
the short-range interaction strength $\eps$, and can be discarded
in the limit of strong short-range repulsion. The salt
term, in turn, pushes the spinodal line downwards, as was previously
shown by Ref.~\cite{harries1,harries2}. The  decrease in the spinodal line
is substantial, and depends on the amount of added salt. In this
(Debye H\"{u}ckel) approximation, for very large amounts of added
salt all electrostatic interactions between the monomers and the
surfactants are totally screened, resulting in a strong decrease in $F_{\rm
el}$. The spinodal line then fully recovers the neutral polymer
limit from Ref.~\cite{chatellier}. For lower amounts of salt,
which include most physical cases, the spinodal is pushed
downwards from the pure monolayer spinodal, as seen in Fig.~3a.

%%%%%%%%%%%%
%% FIG 9
\begin{figure}
\includegraphics[keepaspectratio=true,width=105mm,clip=true,trim=60 0 -50 0]{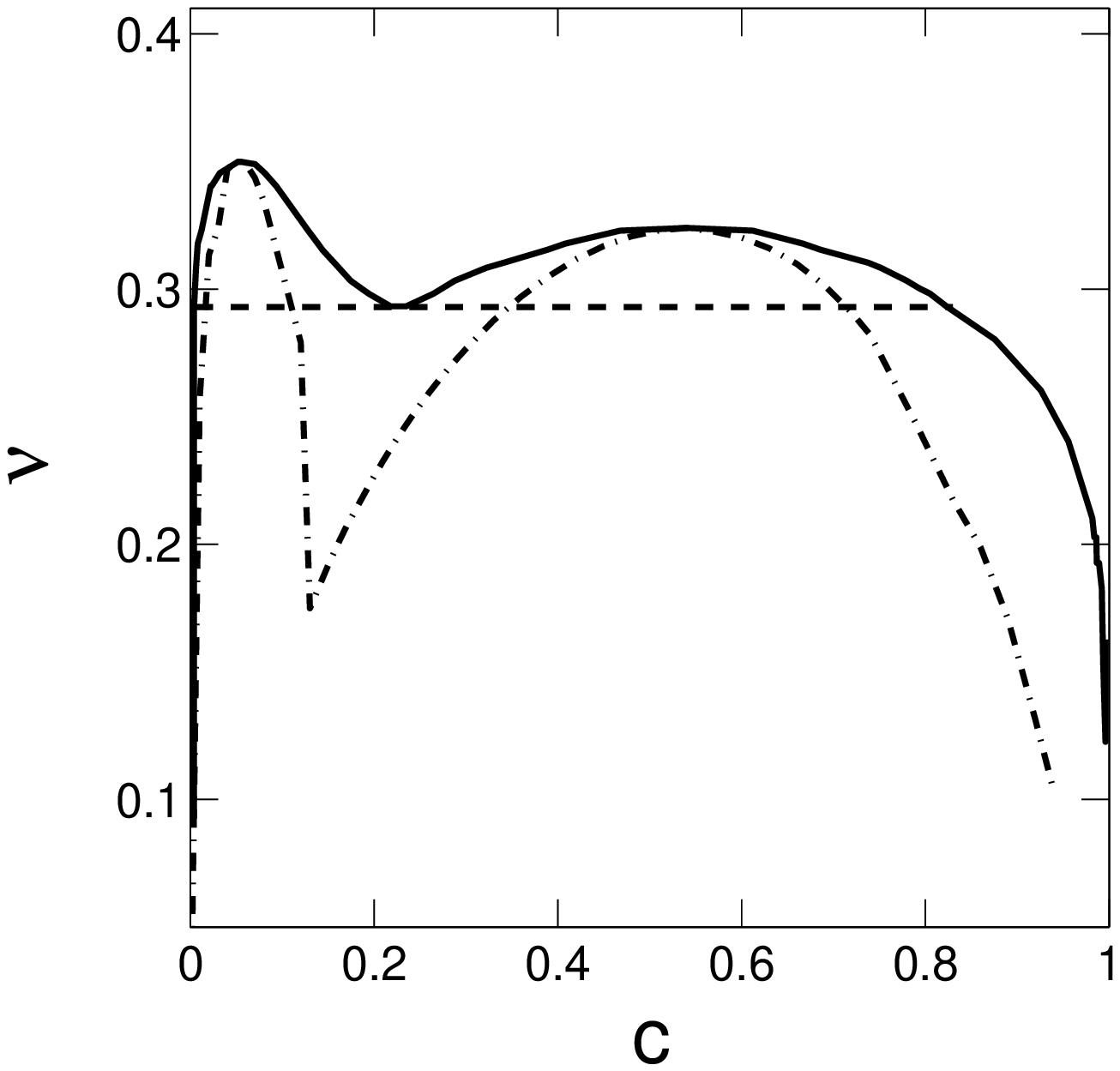}
  \caption{\footnotesize\textsf{
    Binodal and spinodal lines exhibiting a triple point. The dashed-dotted
    line is a numerically calculated spinodal line, for the following
    parameters: $a=5$\,\AA, $b=10$\,\AA, $\nu\cdot\eps=-1.0b^2a$, $f=0.3$,
    $Z=1$, $c^*=0.1$ $\cs=1$\,M, $T=300$\,K, $\phi_b^2=10^{-8}$\,\AA$^{-3}$,
    $v_{\rm ex}=10$\,\AA$^3$. The solid line represents the corresponding
    binodal line. A triple point, connecting three coexisting
    phases, exists for $\nu_{\rm tr}\simeq 0.29$. The three coexisting
    phases are connected by the dashed line.
    }}

\end{figure}

%%%%%%%%%%%%%%%%%%%%%%%%%%%%%%%%%%%%%%%%%%%%%%%%%%%%%%%%%%%%%%%%%%%%%%%%%%%%%%%
\subsection{Electrostatic Adsorption Regime}
\label{pdreps}
%%%%%%%%%%%%%%%%%%%%%%%%%%%%%%%%%%%%%%%%%%%%%%%%%%%%%%%%%%%%%%%%%%%%%%%%%%%%%%%

For $c>c_{\rm dp}$ and $\eps(c-c^*)<0$, the PEs are attracted to the
surface only via their electrostatic interactions, but cannot
approach the surface because of the short-range repulsive
interactions. Due to the exclusion from the surface, the PE surface
concentration becomes very small, and can be ignored at first
approximation.

In previous publications~\cite{itamar1,us1,us2}, we discussed at
length the adsorption of polyelectrolytes to charged surfaces, which
have short-range repulsive interactions with the PE chains. We
summarize here the main results. It was shown that the dominant
interaction is the electrostatic interaction between the PE and the
surface. The excluded volume interaction was found to be negligible
in the numerical solutions of the mean-field
Eqs.~(\ref{normPBmod},\ref{normEdwardsmod}). The PE adsorption was
shown to decrease with the addition of salt~\cite{us1,us2} for low
amounts of added salt, while for higher amounts the polymer chains
were shown to desorb entirely from the surface~\cite{us1}. It was
shown that the adsorbed polyelectrolyte layer can be characterized
by a single (rescaled) monomer concentration scale (denoted by
$\eta_M^2$) and a single length scale for the adsorbed layer width
(denoted by $D$). The free energy of the solution can be calculated
using a Flory-like approximation~\cite{itamar1}. While the full
details can be found in the previous publications, we present here
the main results:

\begin{eqnarray}
 \label{scaledF}
   \frac{F_{\rm ads}}{k_BT}=A_1\frac{a^2}{6}\frac{\phi_b^2\eta_M^2}{D}-
    A_2 f\scharge l_B D^2 \phi_b^2\eta_M^2+ \\ \nonumber
    A_3f^2l_B\phi_b^4\eta_M^4D^3+A_4 \scharge^2 l_B D
\end{eqnarray}
where the surface charge is denoted by $\sigma=-cZ/b^2$.
The first term is the elasticity term of the polyelectrolyte, the
second is the electrostatic attraction between the surface and the monomers,
and the third is the electrostatic repulsion between monomers. This
repulsion is assumed to be strong enough to dominate over the excluded
volume interactions. The last term is the interaction of the electrostatic
potential and the surface charge, where the rescaled potential
is $\ys= A_4\scharge l_B D$. The short-range interaction between
the surfactants and the PE chains depends on the surface monomer
concentration, and can be ignored because of the strong short-range
repulsion. Minimization of Eq.~(\ref{scaledF}) with respect to
$D$ and $\eta_M^2$ yields:

\begin{eqnarray}
   D & \simeq & \left(\frac{a^2}{l_B f \sigma}\right)^{1/3}
 \label{DeqRep} \\
   \phi_b^2 \eta_M^2 & \simeq &  \left( \frac{ \sigma^2 \sqrt{ l_B } }
       { a f }  \right)^{2/3}
 \label{cmeqRep} \\
    F_{\rm ads} & \simeq & I_0
    \left(\frac{a^2l_B^2}{f}\right)^{1/3}\scharge^{5/3}
 \label{FeqRep}
\end{eqnarray}
where the prefactor $I_0$ can be shown to be positive. In this case,
the full free energy can be written as:

\begin{eqnarray}
 \label{Frepphase2}
  F & = & I_0 \left(\frac{a^2l_B^2Z^5}{fb^{10}}\right)^{1/3}c^{5/3}
	 \\ \nonumber & + &
	b^{-2} \bigg[c\ln
  c+\left(1-c\right)\ln\left(1-c\right)+\nu^{-1}c\left(1-c\right)\bigg].   
\end{eqnarray}
Note that this free energy has a fractional power (non-analytic) in $c$.
The spinodal equation is obtained by differentiating
Eq.~(\ref{Frepphase2}) twice with respect to $c$, yielding:

\begin{equation}
  2\nu_{\rm sp}^{-1}=\frac{1}{c}+\frac{1}{1-c}+\frac{10}{9}I_0
  \left(\frac{a^2l_B^2Z^5}{fb^4}\right)^{1/3}c^{-1/3}
 \label{spinRep}
\end{equation}
Eq.~(\ref{spinRep}) shows that the spinodal temperature $\nu_{\rm
sp}$ is always lower than that of the pure
surfactant monolayer, Eq.~(\ref{surfspino}). This result is consistent
with results derived for the spinodals of protein solutions~\cite{may,harries1}.
The critical point is the maximum of this spinodal line $\nu_{\rm
sp}(c)$ in Eq.~(\ref{spinRep}), satisfying:

\begin{equation}
  \frac{2c_c-1}{c_c^2\left(1-c_c\right)^2}=\frac{10}{27}
    I_0  \left(\frac{a^2l_B^2Z^5}{fb^4}\right)^{1/3}c_c^{-4/3}.
 \label{trinRep}
\end{equation}
The critical point is at $c>0.5$, regardless of the sign of $\eps$.
For our chosen parameters of $a=5$\,\AA, $I_0=2.8$, $l_B\simeq 7$\,\AA,
$b=10$\,\AA\, and $Z=1$ the critical concentration from
Eq.~(\ref{trinRep}) turns out to be at $c=0.5365$, showing that the
spinodal is almost symmetric, in agreement with Fig.~3.

%%%%%%%%%%%%%%%%%%%%%%%%%%%%%%%%%%%%%%%%%%%%%%%%%%%%%%%%%%%%%%%%%%%%%%%%%%%%%%
\subsection{Strong Short-Range Attraction Regime}
\label{pdAtt}
%%%%%%%%%%%%%%%%%%%%%%%%%%%%%%%%%%%%%%%%%%%%%%%%%%%%%%%%%%%%%%%%%%%%%%%%%%%%%%

For large enough $\eps(c-c^*)>0$ (strong short range attraction)
the adsorbing polymer layer is highly charged,
more than the initial surface~\cite{wang1,us2}. In this case, the
main electrostatic contribution to the free energy comes from the
monomer-monomer repulsion, rather than the electrostatic
attraction to the surface.

When the amount of small ions in the solution is small, the
length scale for electrostatic potential decay is the same as that
of the monomer concentration. The single length
scale allows us to use similar scaling ideas as in the repulsive
surface case.  The highest monomer concentration occurs at the
surface itself $\eta_{\rm max}=\eta_s$, and acts as the scale for the
monomer concentration. Similarly to Ref.~\cite{itamar1}, we can
produce scaling rules for the monomer adsorption using a
Flory-like  free energy. The free energy in this case
is:

\begin{eqnarray}
 \label{scaledFAtt}
   \frac{F_{\rm ads}}{k_BT}=A_1\frac{a^2}{6}\frac{\phi_b^2\eta_s^2}{D}-
    \half b^{-2}\eps\left(c-c^*\right)\phi_b^2\eta_s^2
     \\ \nonumber+A_3f^2l_B\phi_b^4\eta_s^4D^3
\end{eqnarray}
where the first term is again the contribution of the chain elasticity,
the second term is the short-range attraction of the polymer chains to
the surface, and the last term is the electrostatic repulsion between the
monomer layers. The electrostatic interactions between the surface
and the monomer concentrations are assumed to be small in comparison
to the short-range ones, and are omitted from this free energy.
Minimization of this free energy yields the following
scales for $D$ and $\eta_s^2$:

\begin{eqnarray}
  D & \simeq & \frac{a^2b^2}{\eps\left(c-c^*\right)}
 \label{DeqAtt} \\
  \eta_s^2 & \simeq & \frac{\eps^4\left(c-c^*\right)^4}{b^8a^6l_Bf^2\phi_b^2}
 \label{cmeqAtt} \\
  \frac{F_{\rm ads}}{k_BT} & \simeq & -I_1 \frac{\eps^5\left(c-c^*\right)^5}{b^{10}a^6l_Bf^2}
 \label{FeqAtt1}
\end{eqnarray}
where $I_1$ is a constant of order unity. The adsorption free energy
is negative, as expected, and is very strongly dependent on the
strength of the short range interactions. The total free energy is
now the sum of the pure monolayer free energy and the adsorption
free energy, and the spinodal line can be readily obtained as:

\begin{equation}
  \frac{1}{c\left(1-c\right)}-2\nu_{\rm sp}^{-1}-
  20I_1\frac{\Theta^5\left(c-c^*\right)^3}
    {\nu_{\rm sp}^5l_Ba^6b^8f^2}=0
 \label{SpinAtt}
\end{equation}
where we denote $\Theta\equiv \nu\cdot\eps$. In this case, the
spinodal temperature $\nu_{\rm sp}$ is always larger than the
corresponding temperature of the pure-surface spinodal. The
critical point is obtained by differentiating Eq.~(\ref{SpinAtt})
once more with respect to $c$.

\begin{equation}
  -\frac{1}{c_c^2}+\frac{1}{\left(1-c_c\right)^2}-
  60I_1\frac{\Theta^5\left(c_c-c^*\right)^2}{\nu_c^5l_Ba^6b^8f^2}
    =0
 \label{CritPointAtt}
\end{equation}
which for $\Theta>0$ gives $c_c>0.5$, and vice versa. The
resulting spinodal line is less symmetric than in the

\begin{widetext}

%%%%%%%%%%%%
% FIG 10
\begin{figure}
\includegraphics[keepaspectratio=true,width=190mm,clip=true,trim=30 120 -30 150]{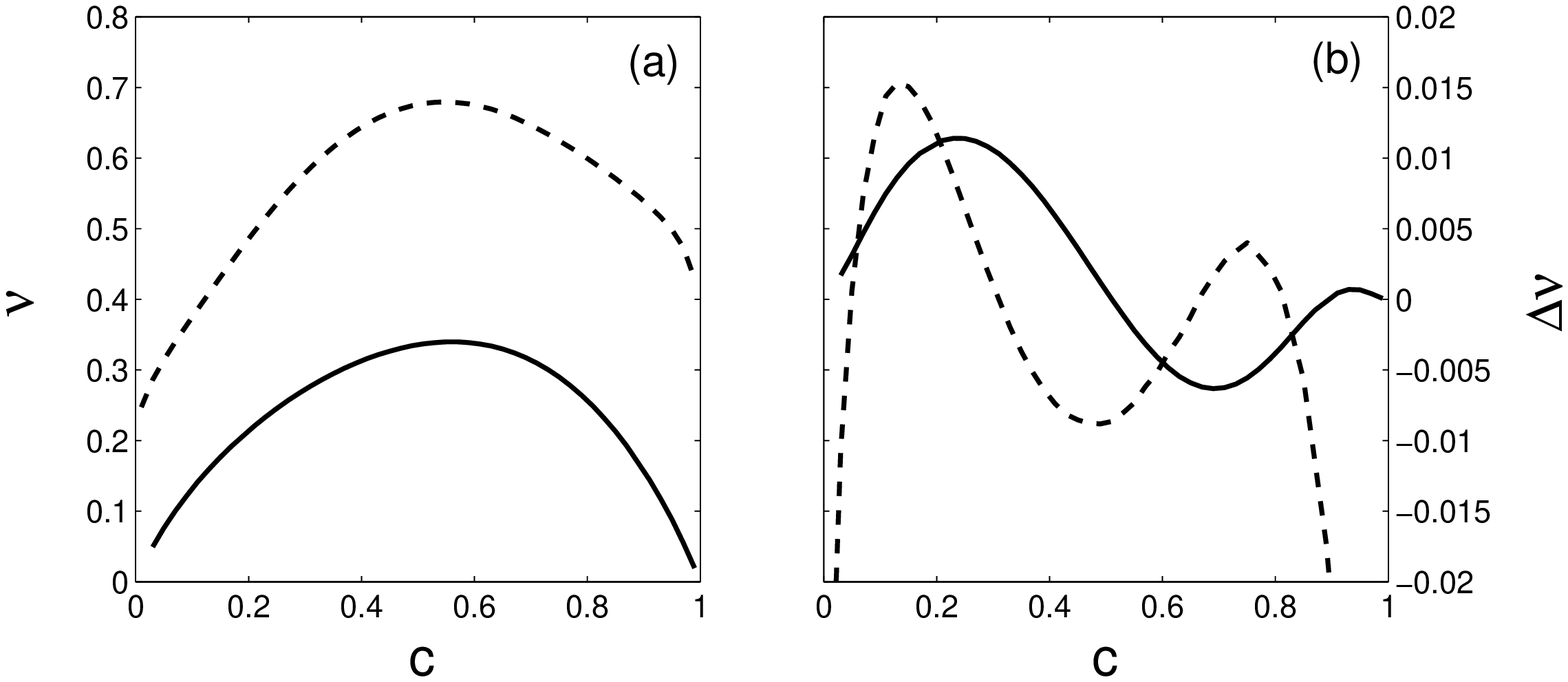}
  \caption{\footnotesize\textsf{
    Comparison between the numerical and analytical spinodal
    lines. In (a) we present the numerical spinodal lines for two sets
    of parameters. The solid line is the numerical
    spinodal line in the repulsive case for $Z=1,\, c^*=3.0$, while the
    dashed line is the
    numerical spinodal line in the attractive case, for $Z=0.1, \, c^*=-1.0$.
    Both spinodal
    lines share $a=5$\,\AA, $b=10$\,\AA, $\nu\cdot\eps=0.5b^2a$, $f=1$,
    $\cs=0.01$\,M, $T=300$\,K, $\phi_b^2=10^{-8}$\,\AA$^{-3}$,
    $v_{\rm ex}=10$\,\AA$^3$. We do not show the analytical spinodals
    here since they are too close to the numerical ones to be drawn.
    In (b) we present the differences between the
    analytical calculations and the numerical spinodals.
    The solid and dashed lines are the difference between
    the spinodal lines in Eqs.~(\ref{spinRep}) and (\ref{SpinAtt}),
    respectively, with $I_0=2.8$ and  $I_1=0.023$, and the numerical
    spinodals in (a). The differences are small, and the
    numerical spinodal lines are well characterized by the analytical
    approximation.
}}
\end{figure}
\end{widetext}
electrostatic adsorption case, and the critical point is further
from the pure surface value of $0.5$ than the repulsive-surface
spinodal Eq~(\ref{spinRep}). This result is also in agreement
with the numerical results of Fig.~5.

%%%%%%%%%%%%%%%%%%%%%%%%%%%%%%%%%%%%%%%%%%%%%%%%%%%%%%%%%%%%%%%%%%%%%%%%%%
\subsection{Comparison of the analytical and numerical results}
\label{CompRes}
%%%%%%%%%%%%%%%%%%%%%%%%%%%%%%%%%%%%%%%%%%%%%%%%%%%%%%%%%%%%%%%%%%%%%%%%%%

Both spinodals, Eqs.~(\ref{spinRep},\ref{SpinAtt}), contain one
free parameter each, and can be compared to the numerical
spinodals. The comparison is presented in Fig.~10. As can be seen, both
of the spinodal equations show good agreement with the
numerical lines, using the values $I_0=2.8$ for the repulsive
surface case and $I_1=0.023$ for the attractive surface case.

The analytical model given above for the two adsorption regimes is
accurate only for the case of low amounts of added salt. This is
because the main assumption is the existence of a single length
scale for the decay of the electrostatic potential and the
monomer concentration. More work is necessary to extend the
present model for the high salt regime.

%%%%%%%%%%%%%%%%%%%%%%%%%%%%%%%%%%%%%%%%%%%%%%%%%%%%%%%%%%%%%%%%%%%%%%%%%%%%
%%%%%%%%%%%%%%%%%%%%%%%%%%%%%%%%%%%%%%%%%%%%%%%%%%%%%%%%%%%%%%%%%%%%%%%%%%%%
\section{Conclusions}
\label{conclusion}
%%%%%%%%%%%%%%%%%%%%%%%%%%%%%%%%%%%%%%%%%%%%%%%%%%%%%%%%%%%%%%%%%%%%%%%%%%%
%%%%%%%%%%%%%%%%%%%%%%%%%%%%%%%%%%%%%%%%%%%%%%%%%%%%%%%%%%%%%%%%%%%%%%%%%%%

We present numerical and analytical results for the phase separation of
a surfactant monolayer in presence of polyelectrolytes (PEs). We show that
the short-range interactions between the surfactant and polyelectrolytes
have a strong impact on the surfactant critical temperature and critical
concentration. We also present analytical results in two
limits: (i) strong short-range repulsive, and (ii) attractive
surfactant-polyelectrolyte interactions.

For short-range repulsive interactions, our results show that
surfactant phase separation occurs for lower temperatures than for a
non-interacting surfactant layer. For such systems, the
polyelectrolyte chains may desorb completely from the low surfactant
concentration domains, while for higher concentration phases the PE
adsorption increase linearly with the surfactant surface coverage.
The critical surfactant concentration slightly deviates from the
bare one $c_c=0.5$, showing nearly symmetrical spinodal and binodal
lines. In addition, we find a temperature range in which phase
separation does not increase the PE adsorption. For lower
temperatures, where the phase separation occurs between a strongly
adsorbing phase and a depleting phase, the gain in adsorption
becomes substantial. For short-range attractive interactions, in
contrast, the phase separation always increases adsorption, and the
demixing temperatures are always larger than the pure-surface ones.
The critical point moves more significantly away from the
pure-surface symmetric value. The convexity of the adsorption is
more pronounced, showing that phase separation indeed increases the
PE adsorption on the surfactant monolayer. In intermediate cases,
where the short-range interaction between the polyelectrolyte chains
and the surface changes sign as function of the surfactant surface
coverages, a triple point can be found on the binodal line
connecting three coexisting phases on the same surface.

The addition of salt is shown to increase the threshold
surfactant-concentration for the desorption phase. It has a weaker
effect on the spinodal line for the case of electrostatic adsorption.
For short-range adsorbing surfaces, the addition of salt
increases the spinodal line considerably.

Our results are derived within mean-field, using the ground state
dominance approximation. This treatment neglects correlation
effects between the different constituents, and they may
be strong in some cases. We also assume that the lipids are entirely
insoluble, and neglect the effects of the line tension between the
surfactant phases. Despite these drawbacks, we believe that our
main results still hold. Our results may serve as a
starting point for more complex models.

\vskip 2truecm

{\it Acknowledgments}: We thank Avinoam Ben-Shaul, Yoram Burak and Shelly
Tzlil for helpful discussions. Support from the Israel Science Foundation
(ISF) under grant no.
160/05 and the US-Israel Binational Foundation (BSF) under grant no.
287/02 is gratefully acknowledged.

%%%%%%%%%%%%%%%%%%%%%%%%%%%%%%%%%%%%%%%%%%%%%%%%%%%%%%%%%%%%%%%%%%%%%%%%%%%%
%\newpage

%%%%%%%%%%%%%%%%%%%%%%%%%%%%%%%%%%%%%%%%%%%%%%%%%%%%%%%%%%%%%%%%

\end{document}